\documentclass[a4paper,11pt]{article}
\usepackage{pos}
\usepackage{subcaption}
\title{Nuclear PDFs}

\author*[a]{K. Greve}
\author[a]{T. Je\v{z}o}
\author[a]{M. Klasen}
\author[b]{A. Kusina}
\author[c]{F.I. Olness}
\author[a]{J. Wissmann}
\affiliation[a]{Institute for Theoretical Physics, University of M{\"u}nster, Wilhelm-Klemm-Stra{\ss}e 9, 48149 M{\"u}nster, Germany}
\affiliation[b]{Institute of Nuclear Physics, Polish Academy of Sciences, 31342 Krakow, Poland}
\affiliation[c]{Department of Physics, Southern Methodist University, Dallas, TX 75275-0175, U.S.A.}
\emailAdd{kgreve@uni-muenster.de}
\emailAdd{tomas.jezo@uni-muenster.de}
\emailAdd{aleksander.Kusina@ifj.edu.pl}
\emailAdd{michael.klasen@uni-muenster.de}
\emailAdd{olness@smu.edu}
\emailAdd{jan.wissmann@uni-muenster.de}

\abstract{We summarize the current status of global analyses of model-independent nuclear parton distribution functions (nPDFs). 
A carefully conducted global analysis of nPDFs in the era of heavy-ion collisions at the LHC is key to gaining access to new information through the data measured by the collider experiments.
This implies incorporating as large and diverse a dataset as possible, 
as well as the latest theoretical calculations, in order to 
capture a faithful representation of nuclear structure
that reflects the current state of knowledge. 
On the other hand, the uncertainties in particular must reflect what is not known.
For many observables, nuclear uncertainties represent the most significant source of error in current theoretical models; nevertheless, they should not, where possible, be artificially constrained (for example, through parameterization bias).
The quality of a global (n)PDF fit is often measured by the $\chi^2$ per degree of freedom, which is approximately one for a 
fit that describes the data precisely without overfitting.
Many additional aspects allow to characterize a good fit, e.g. the $\chi^2$ distribution per experiment, the shape of the uncertainties, and the sensitivity to different types of data.
We inspect various fit quality indicators for fits obtained by different groups.
Morever we pay close attention to the $A$-dependence.
Traditionally, the $A$-dependence had a primarily functional role in global nPDF analyses and comparatively less attention has been paid to it when characterizing a fit.
However, we have now entered the era of nuclei collisions at the LHC and need to investigate the $A$-dependence properly.
}

\FullConference{14th Edition of the Large Hadron Collider Physics (LHCP2026)\\
18-22 May 2026\\
Paris, France\\}

\begin{document}
\maketitle

\section{Introduction}
These proceedings compare, in particular, the EPPS21 \cite{EPPS21} and nNNPDF3.0 \cite{nNNPDF30} analyses with preliminary results of the nCTEQ26 analysis.
All three analyses are carried out with next-to-leading (NLO) precision.
Additionally, in the first section we show a comparison with the TUJU21 global analysis, carried out at NNLO \cite{TUJU21}.
A comprehensive review of nPDFs after the first decade of the LHC is presented in Ref. \cite{review_Nuclear_PDF}. 

\section{Experimental Data}
\begin{figure}
  \centering
  \subfloat[nCTEQ26 (prel.)]{
  \includegraphics[width=0.38\linewidth]{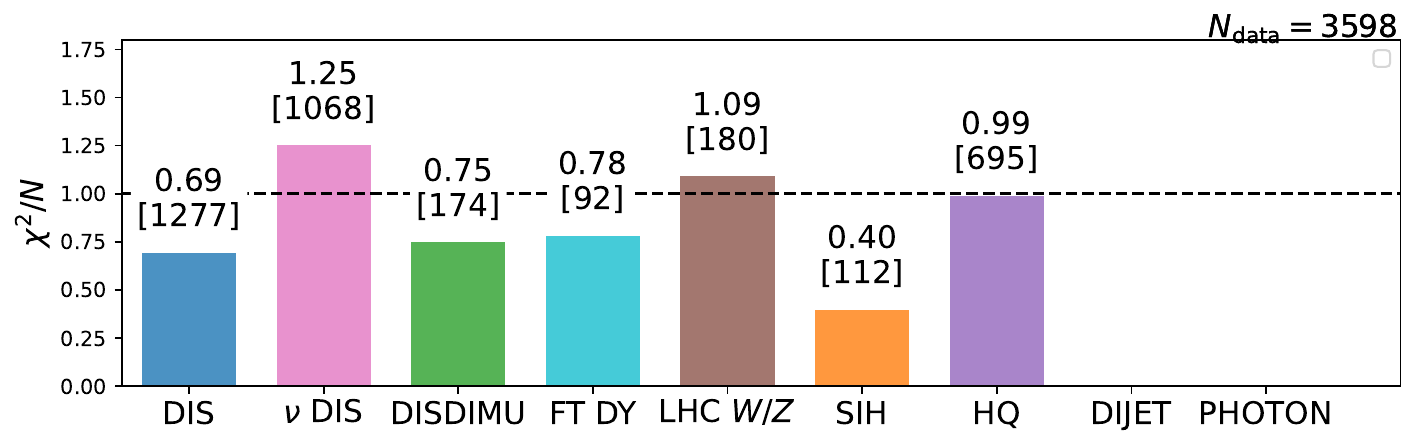}
  }
  \subfloat[EPPS21]{
  \includegraphics[width=0.38\linewidth]{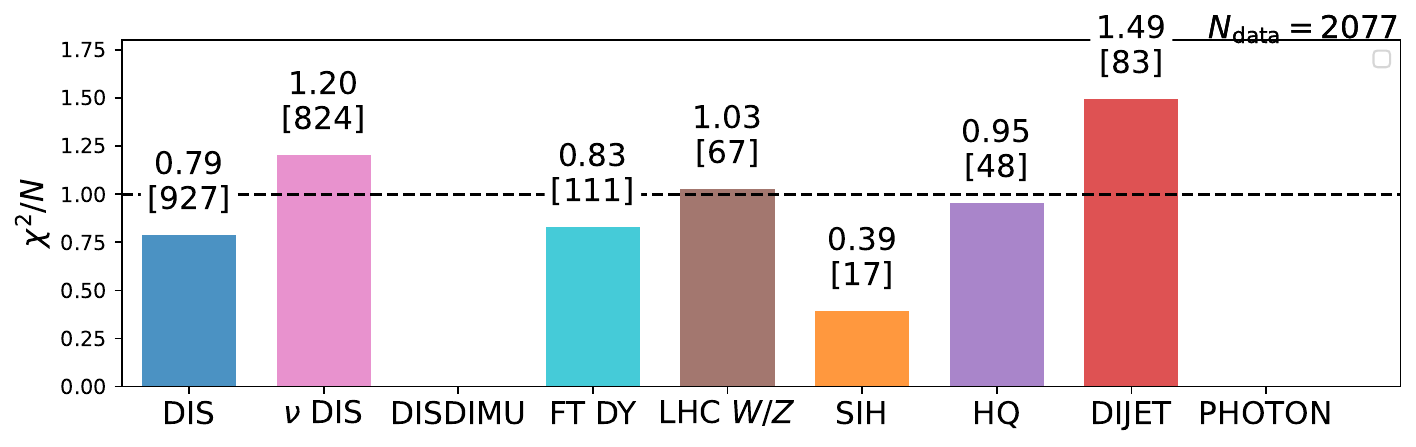}
  }
  \\
  \subfloat[nNNPDF3.0]{
  \includegraphics[width=0.38\linewidth]{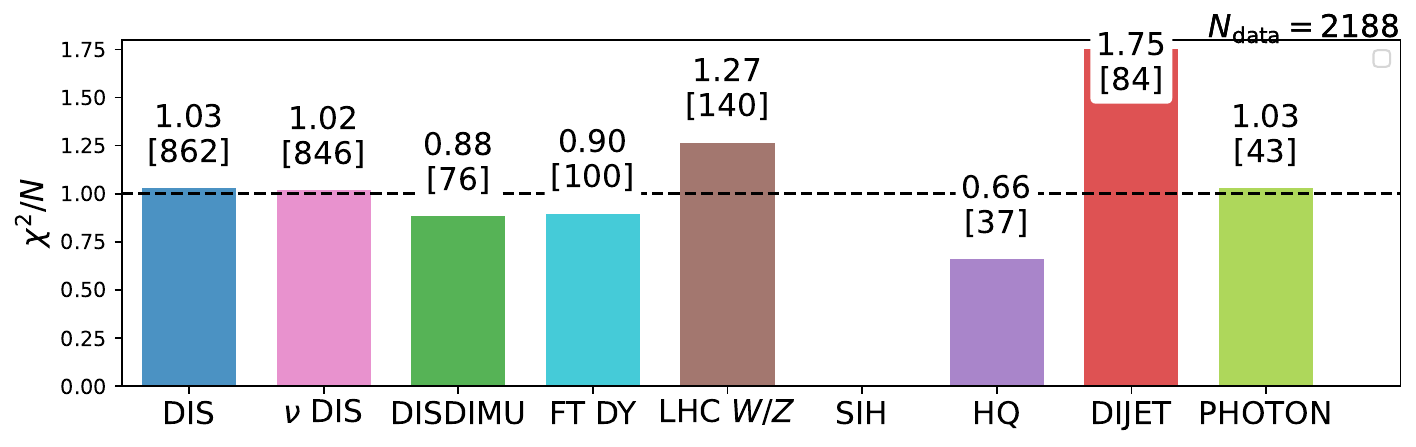}}
  \subfloat[TUJU21]{
  \includegraphics[width=0.38\linewidth]{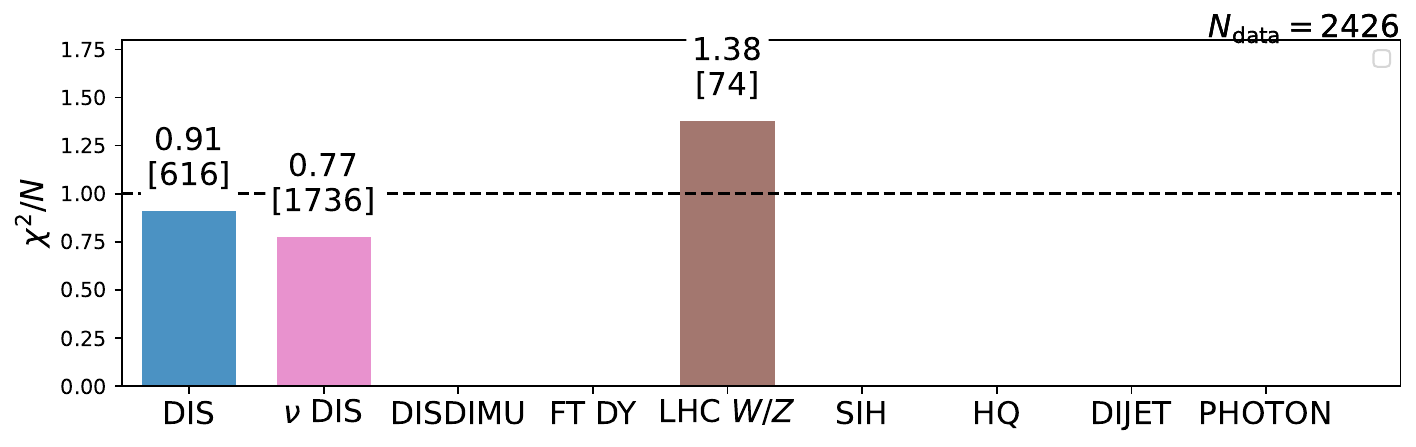}
  }
  \caption{Overview on data and $\chi^2/n_\text{data}$ per type of experiment for different model independent nPDF analyses. This data was manually extracted from the corresponding publications.}
  \label{fig:bar_data}
\end{figure}

The majority of the experimental data used in nPDF analyses ($\sim 10^3$~points) come from deep inelastic scattering (DIS) experiments, which provide direct access to the structure functions of the nucleus.
A distinction is made here between neutral-current, charged-current (neutrino scattering, $\nu$~DIS) and dimuon production (neutrino scattering with two muons in the final state, DISDIMU),
wich provide access to different linear combinations of PDFs. 
In fixed-target Drell Yan (FT DY) dileptons are produced by two hadrons in the initial state.
Hadron-hadron collider data, that is typically considered in nPDF analyses ($\sim 10^2$~points), come from the LHC in form of $W/Z$ production experiments, single-inclusive-hadron (SIH) production, heavy-quark (HQ) production, dijet measurements and prompt photons.
nCTEQ26 includes an especially large amount of HQ data.
The bar charts of Fig.~\ref{fig:bar_data} show the $\chi^2$ per type of experiment for the NLO analyses nCTEQ26 preliminary, EPPS21 and nNNPDF3.0
as well as for the NNLO TUJU21 analysis.
Overall the DIS, dimuons, FT DY and HQ data are well described by all analyses.
The agreement with LHC $W/Z$ and $\nu$ DIS data is a bit weaker.
The CMS dijets (only included by EPPS21 and nNNPDF3.0) are generally challenging to describe. 
It is worth emphasizing the excellent agreement between nCTEQ26 and EPPS21 across both the more straightforward data sets and those that are more challenging to describe theoretically.
TUJU21 includes only neutral and charged-current DIS data and LHC $W/Z$ production data. 
Therefore it can not 
be considered as comprehensive as the other global analyses.
Nevertheless, as one of the few nPDF analyses at the NNLO level, this work provides valuable guidance for the future of nPDF research.
Another aspect worth mentioning is the conceptually different classification of the deuteron. 
nNNPDF3.0 and TUJU21 classify the deuteron as a nucleus distinct from the proton and include the data in their analyses. 
EPPS21 and nCTEQ26 use proton baselines that have already taken into account the deuteron data with corrections to it and therefore do not include them in the nuclear fit to avoid double counting.

\section{Comparison of PDFs}

\begin{figure}
  \centering
  \includegraphics[width=0.6\linewidth]{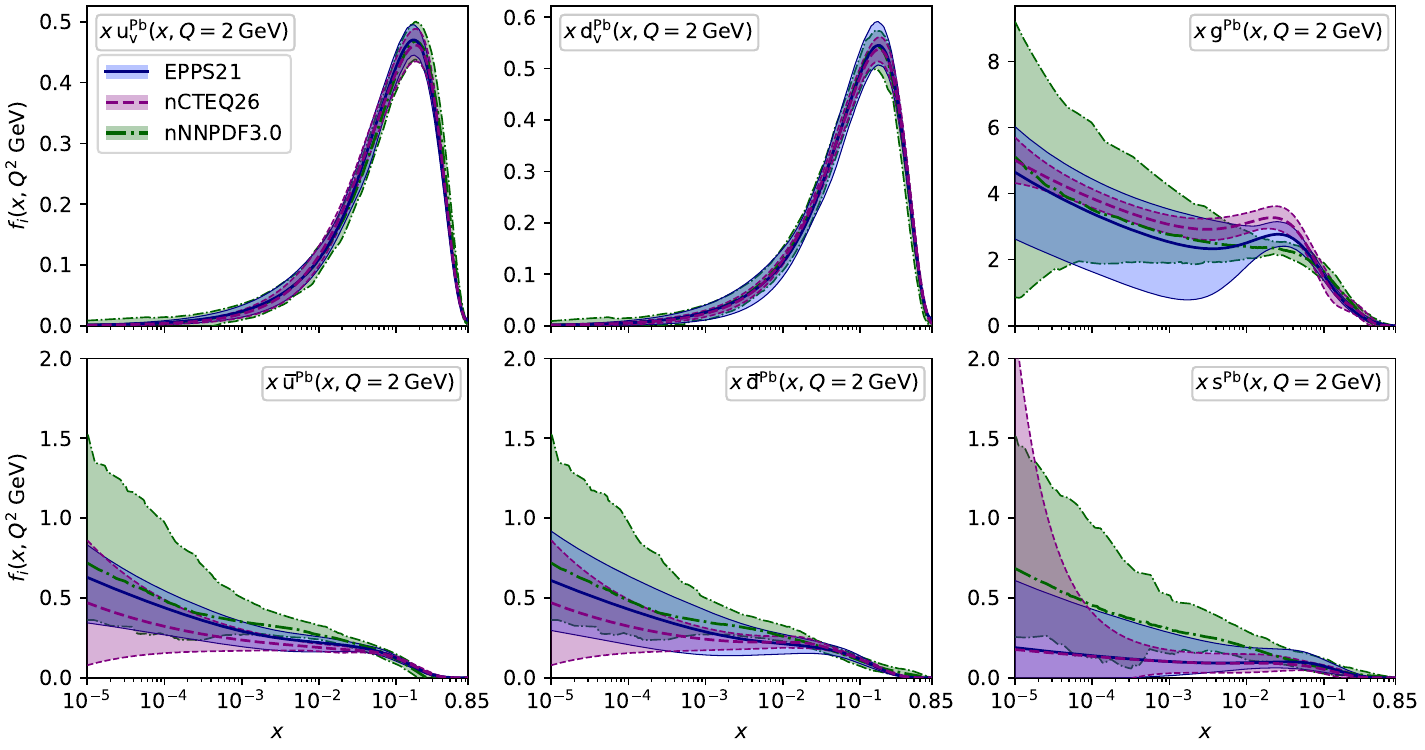}
  \caption{Pb nuclear PDFs for the presented analyses at NLO theory precision at $Q=2$ GeV. The error bands represent the 90\% confidence levels.} 
  \label{fig:pdfs}
\end{figure}

In Fig. \ref{fig:pdfs} the discussed NLO nPDFs with 90\% confidence level uncertainty bands for several flavors for lead ($A=208$) at $Q=2$ GeV are presented.
nCTEQ26 and EPPS21 calculate their uncertainties using the Hessian method, and nNNPDF3.0 uses Monte Carlo replicas. 
It is worth mentioning that there are first studies on the uncertainty estimation using Monte Carlo Markov Chains, see Ref. \cite{derakhshanianDeterminationNuclearPDFs2026}.
In the upper left and upper center plots of Fig. \ref{fig:pdfs}, one can see the $u$-valence and $d$-valence nPDFs. 
One can clearly observe a very good agreement between all collaborations, in the central values as well as in the size of uncertainty bands.  
This agreement is significant because the analyses use different combinations of experimental data, theoretical inputs, parameterizations, and fitting methodologies. 
Their convergence therefore provides important validation of the global PDF framework and demonstrates the robustness of the extracted nPDFs despite these independent methodological and data-driven differences.
The situation for the gluon distribution, upper right, is a bit different. 
Here, nCTEQ26 has clearly the smallest uncertainty band, especially at low $x$. 
This can be explained by the already mentioned large amount of HQ data included in the analysis.
In the lower row, the sea quark distributions are shown. 
nNNPDF3.0 has similar values for $\overline{u}$, $\overline{d}$ and strange PDF (a democratic sea).
Both EPPS21 and nCTEQ26 have a suppressed strange quark distribution, which is consistent with the suppressed gluon to strange splitting due to the strange mass as observed in fixed-target neutrino DIS.

\section{Nuclear modifications and evidence for nuclear effects}
\begin{figure}
  \centering
  \subfloat[Nuclear modifications for lead ($A=208)$ for different flavors at $Q=2$ GeV.]{
  \centering
  \includegraphics[width=0.6\linewidth]{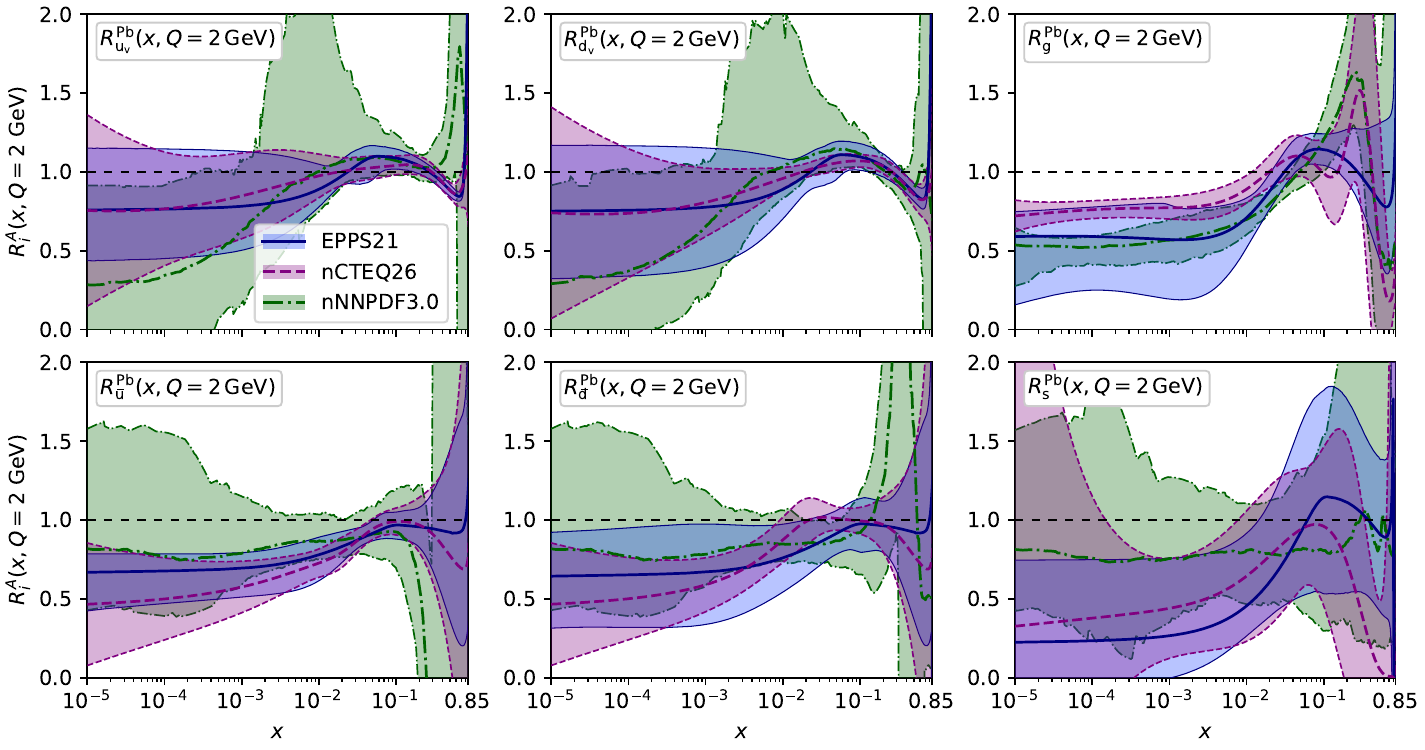}}
  \\
  \subfloat[Nuclear modifications for lead ($A=208)$ for different flavors at $Q=10$~GeV.]{
  \centering
  \includegraphics[width=0.6\linewidth]{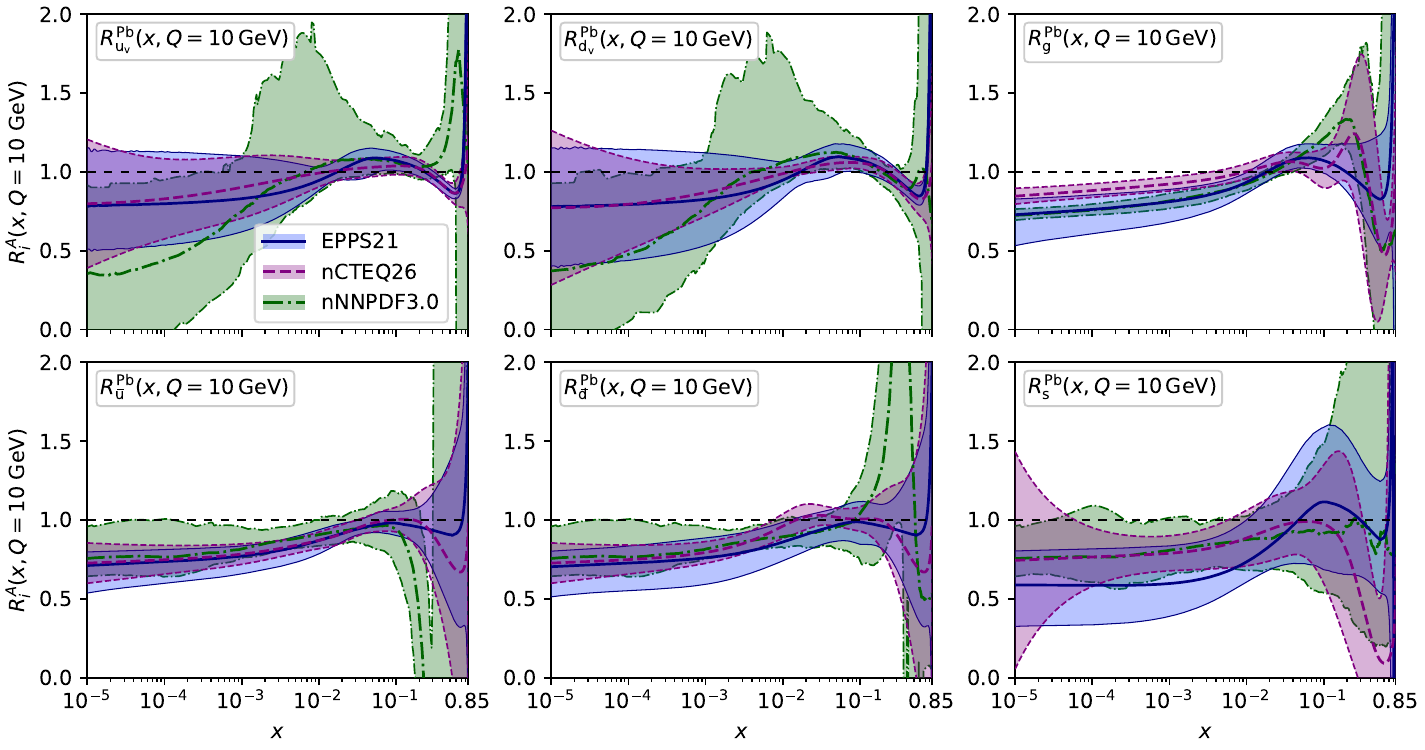}}
  \caption{Nuclear modifications for lead (A=208) for the three NLO global nPDF analyses nCTEQ26 (magenta), EPPS21 (blue) and nNNPDF3.0 (green).}
  \label{fig:nuc_mod_lead}
\end{figure}

Nuclear effects, i.e.~the difference between the free proton PDF and the bound proton PDF inside a nucleus, can be quantified by the nuclear modification ratio 
\begin{align}
   R_i^A(x, Q^2) = \frac{Z f_i^{(p/A)}(x, Q^2)+(A-Z) f_i^{(n/A)}(x, Q^2)}{Z f_i^{(p)}(x, Q^2)+(A-Z) f_i^{(n)}(x, Q^2)},
\end{align}
such that a ratio equal to unity means there is no nuclear effect.
In Fig.~\ref{fig:nuc_mod_lead}, the nuclear modifications for lead are shown for the three presented NLO analyses.
In the upper plot the modifications at $Q=2$~GeV are shown and in the lower one for $Q=10$~GeV.
At the lower energy the most pronounced nuclear effect is the gluon shadowing at low $x$.
All three groups also found shadowing for $u_v$ and $d_v$ as well as antishadowing for $u_v$, $d_v$ and gluon but with an uncertainty a bit below/above unity.
EPPS21 and nCTEQ26 also show deviations from unity within the uncertainties in the shadowing regions of the $\overline{u}$ and $\overline{d}$ and strange modifications.
Comparing the overall picture to the situation in Fig. \ref{fig:nuc_mod_lead} b) where the modifications are shown at $Q=10$~GeV, one can see that there is a general trend for modifications being weaker (meaning central values closer to one) at higher energies but more significant (meaning smaller uncertainties).

\section{$A$-dependence}
\begin{figure}
  \centering
  \includegraphics[width= 0.76\linewidth]{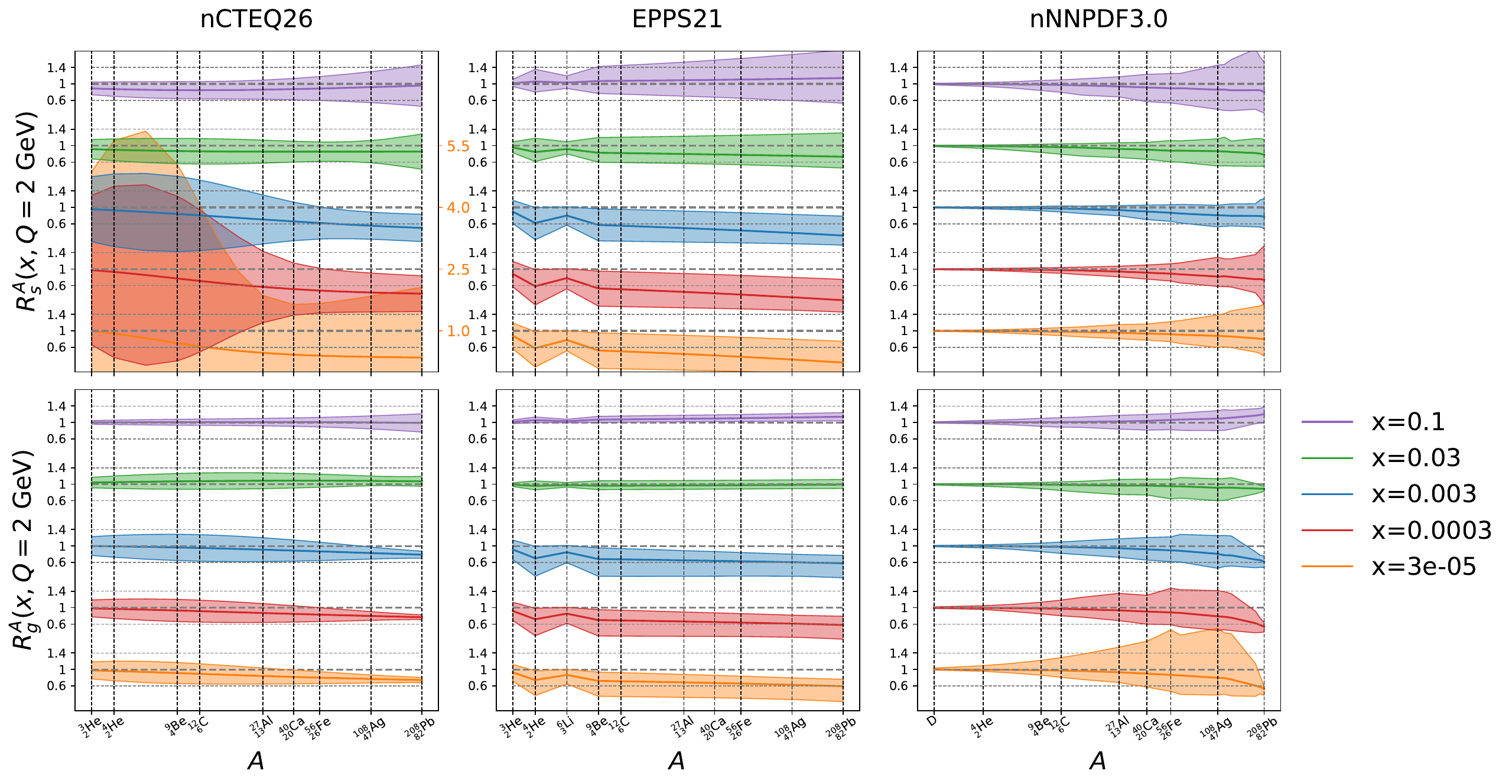}
  \caption{$A$-dependence of nuclear modifications at $Q=2$ GeV for different values of $x$.}
  \label{fig:adep}
\end{figure}

To investigate the $A$ dependence in more detail, Fig.~\ref{fig:adep} is presented.
The upper row shows the strange modifications for the three analyses, the lower row the gluon modifications.
On the $x$-axis we have the nuclear number $A$; different values of Bjorken $x$ are denoted by different colors and plotted in slices, from low $x$
at the bottom to high $x$
at the top.
The evolution in $A$ of the uncertainties for strange looks quite different for the different analyses. 
nCTEQ26 has a minimum width around Fe for low and high $x$. This can be explained by the inclusion of NuTeV/CCFR dimuon iron scattering data, which is the only data directly sensitive to strange.
The strange uncertainty for EPPS21 is slightly increasing towards heavy nuclei.
The same effect, but more drastic, is seen for nNNPDF3.0.
Here, the free-proton boundary condition for light nuclei, as well as the inclusion of deuteron data, leads to small uncertainties for light nuclei.
The situation for the gluon is different.
Here, EPPS21 and nCTEQ26 tend to have decreasing uncertainties towards lead for low $x$ and increasing uncertainties for high $x$, which can be explained by the inclusion of the low-$x$ HQ data from lead collisions at the LHC.
However, this effect is very small for EPPS21 and more pronounced for nCTEQ26.
EPPS21 being relatively flat in $A$ is expected, as the analysis uses only five parameters that can distinguish between different nuclei.
In contrast, nNNPDF3.0 shows a pronounced maximum width around Ag, since for light nuclei there are the constraints discussed above, and for heavy nuclei the lead LHC data can constrain the modification. \
Comparing nCTEQ26 to older nCTEQ results (not shown), e.g. nCTEQ15 \cite{nCTEQ15} or nCTEQ15HQ \cite{nCTEQ15HQ} shows that 
the inclusion of new data does not necessarily lead to smaller uncertainties. 
In particular, if the uncertainties in earlier analyses were underestimated, adding new data can allow to use a more flexible parametrization and therefore instead reveal a larger underlying uncertainty. 
This effect is especially evident for light nuclei.

\section{Summary}
Overall, there is good agreement among the current NLO nPDF analyses.
The different collaborations rely largely on similar experimental data sets
while data for light nuclei remain relatively scarce, particularly for hadron-hadron collisions. 
State of the art nPDFs have relatively large uncertainties and in many kinematic regions the modification ratio of one is included within the 90\% confidence intervalls.
However, the uncertainties in older analyses may have been underestimated, while new data provide stronger and more reliable constraints on nuclear effects in these regions.
The uncertainty landscape in $A$ is one of the main differences between the analyses. 
The distinct uncertainty patterns reflect differences in the included data sets, the treatment of the deuteron, and the chosen A-dependence of the PDF parametrization.

\end{document}